\documentclass[twocolumn,english,aps,prb,showpacs,byrevtex,amsmath,amssymb,bm,superscriptaddress]{revtex4-1}
\usepackage[T1]{fontenc}
\usepackage[latin9]{inputenc}
\usepackage{textcomp}
\usepackage{amsmath}
\usepackage{graphicx}

\makeatletter
\usepackage{lipsum}
\usepackage{epsfig}
\usepackage{color}
\usepackage{physics}

\definecolor{Red}{rgb}{1,0,0}
\definecolor{Blu}{rgb}{0,0,1}
\definecolor{Green}{rgb}{0,1,0}

\makeatother

\usepackage{babel}
\begin{document}

\title{Spin-orbit coupling effects on the electronic properties of the pressure-induced superconductor CrAs}

\author{Giuseppe Cuono}
\email{gcuono@unisa.it}
\affiliation{Dipartimento di Fisica "E.R. Caianiello", Universit\`a degli Studi di Salerno, I-84084 Fisciano
(SA), Italy}

\author{Carmine Autieri}
\affiliation{International Research Centre Magtop, Institute of Physics, Polish Academy of Sciences,
Aleja Lotnik\'ow 32/46, PL-02668 Warsaw, Poland}
\affiliation{Consiglio Nazionale delle Ricerche CNR-SPIN, UOS Salerno, I-84084 Fisciano (Salerno),
Italy}

\author{Giuseppe Guarnaccia}

\affiliation{Dipartimento di Fisica "E.R. Caianiello", Universit\`a degli Studi di Salerno, I-84084 Fisciano
(SA), Italy}

\author{Adolfo Avella}

\affiliation{Dipartimento di Fisica "E.R. Caianiello", Universit\`a degli Studi di Salerno, I-84084 Fisciano
(SA), Italy}

\affiliation{Consiglio Nazionale delle Ricerche CNR-SPIN, UOS Salerno, I-84084 Fisciano (Salerno),
Italy}

\affiliation{Unit\`a CNISM di Salerno, Universit\`a degli Studi di Salerno, I-84084 Fisciano (Salerno), Italy}

\author{Mario Cuoco}

\affiliation{Consiglio Nazionale delle Ricerche CNR-SPIN, UOS Salerno, I-84084 Fisciano (Salerno),
Italy}

\affiliation{Dipartimento di Fisica "E.R. Caianiello", Universit\`a degli Studi di Salerno, I-84084 Fisciano
(SA), Italy}

\author{Filomena Forte}

\affiliation{Consiglio Nazionale delle Ricerche CNR-SPIN, UOS Salerno, I-84084 Fisciano (Salerno),
Italy}

\affiliation{Dipartimento di Fisica "E.R. Caianiello", Universit\`a degli Studi di Salerno, I-84084 Fisciano
(SA), Italy}

\author{Canio Noce}
\affiliation{Dipartimento di Fisica "E.R. Caianiello", Universit\`a degli Studi di Salerno, I-84084 Fisciano
(SA), Italy}
\affiliation{Consiglio Nazionale delle Ricerche CNR-SPIN, UOS Salerno, I-84084 Fisciano (Salerno),
Italy}

\date{\today}
\begin{abstract}
		
We present the effects of spin-orbit coupling on the low-energy bands and Fermi surface of the recently discovered pressure-induced superconductor CrAs. We apply the L\"{o}wdin down-folding procedure to a tight-binding hamiltonian that includes the intrinsic spin-orbit interaction, originating from the Cr 3$d$ electrons as well as from As 4$p$ ones. Our results indicate that As contributions have negligible effects, whereas the modifications to the band structure and the Fermi surface can be mainly ascribed to the Cr contribution. We show that the inclusion of the spin-orbit interaction allows for a selective removal of the band degeneracy due to the crystal symmetries, along specific high symmetry lines. Such release of the band degeneracy naturally determines a reconstruction of the Fermi surface, including the possibility of changing the number of pockets.
\end{abstract}


\maketitle

\section{Introduction}

In the last years, materials that exhibit a superconducting phase in the vicinity of a quantum critical point have attracted great attention since the superconductivity in such compounds usually may be ascribed to an unconventional formation of Cooper pairs via some critical fluctuations~\cite{Goll06,Norman11}.
In these cases, the superconducting transition is often realized when an external parameter, such as chemical or physical pressure, acts suppressing an high temperature ordered state.
This behavior has been ascertained in the heavy fermion materials~\cite{Stewart84,Amato97,Movshovich01,Avella04}, the cuprate superconductors~\cite{VanHarlingen95,Timusk99,Norman05,Lee06,Armitage10,Rice12,Avella14,Keimer15,Avella16,Novelli17,Avella04b,Krivenko05,Avella07,Avella09,Avella13}, the strontium-ruthenate compounds~\cite{Mackenzie03,Ovchinnikov03,Cuoco06a,Cuoco06b,Forte10,Autieri12,Malvestuto13,Autieri14,Granata16} and the iron-pnictide superconductors~\cite{Mazin08,Stewart11}.

The recently discovered chromium-based superconductor CrAs exhibits a dome-shaped phase diagram very similar to these compounds~\cite{Varma99,Noce00,Noce02,Vandermarel03,Jiang09,Shibauchi14,Seo15} when an external pressure is applied~\cite{Wu14,Kotegawa14}.
It belongs to the family of transition-metal pnictides and at ambient pressure shows an orthorhombic MnP-type (B31) structure, with the unit cell parameters given by $a$=5.649 {\AA}, $b$=3.463 {\AA} and $c$=6.2084 {\AA}~\cite{Wu14}. The primitive cell is formed by four Cr and four As atoms, the Cr atoms being located in the center of CrAs$_6$ octahedra, and surrounded by six nearest-neighbour arsenic atoms (see Fig.1).

\begin{figure}
	\centering
	\includegraphics[width=8cm,angle=0]{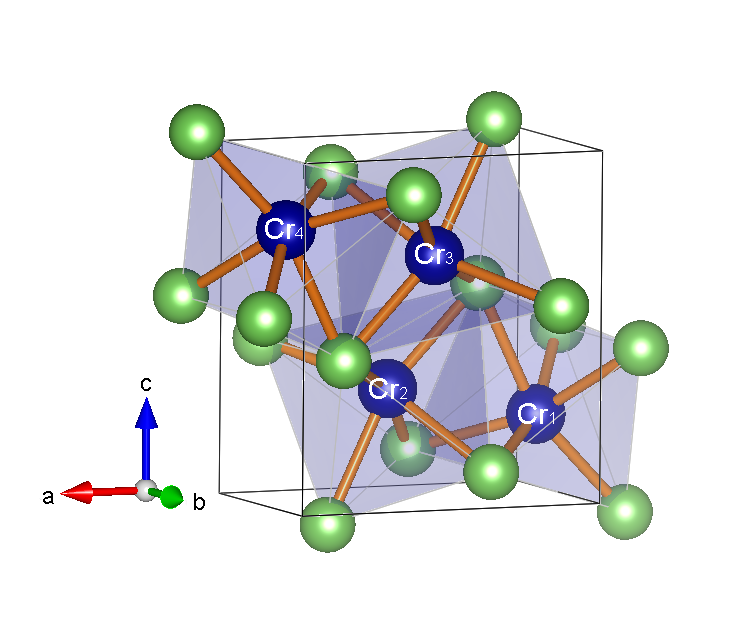}
	\caption{Crystal structure of the CrAs. Cr and As are shown as blue and green spheres, respectively.}
	\label{fig:CrAs}
\end{figure}

Concerning the transport properties, it has been reported a $T^2$ temperature dependence of resistivity, i. e. a Fermi-liquid behavior~\cite{Wu10,Nigro18}, and found a Kadowaki-Woods ratio equal to $1\times10^{-5}\mu \Omega$ cm mol$^2$ K$^2$ mJ$^{-2}$, which fits well to the universal value of many heavy fermion compounds. Moreover, it has been very recently shown that CrAs is a bad metal, differently from Cr, a good metal with a room-temperature resistivity far below the Ioffe-Regel limit~\cite{Matsuda18}.

As far as the magnetic properties are concerned,  at about 270 K a clear magnetic transition  with sharp changes of resistivity and susceptibility has been observed~\cite{Wu10}. Neutron diffraction shows that the CrAs is a non collinear helimagnet and the average ordered magnetic moment is reduced from 1.73 $\mu_B$ at ambient pressure to 0.4 $\mu_B$ close to the critical pressure $P_{c}\sim$ 0.7 GPa, where the magnetic order is completely suppressed~\cite{Keller15}. The magnetic moment direction changes under pressure, with a spin reorientation from the $ab$ plane to the $ac$ plane at the critical pressure~\cite{Shen16}. Furthermore, the helimagnetic order is not only suppressed by the hydrostatic pressure, but also by phosphorus doping in CrAs$_{1-x}$P$_{x}$ at a critical $x_{c}\sim$ 0.05~\cite{Matsuda18}. Interestingly, a study on Al doped CrAs indicates that $T_{N}$ is similar to that of the undoped CrAs, and when an external pressure is applied, $T_{N}$ vanishes near 4.5 kbar, a value shifted from the $P_{c}$ of 9 kbar of undoped CrAs. Nevertheless, the pressure-induced superconductivity is almost independent of Al doping~\cite{Park18}.

The superconductivity appears when the external pressure reaches a critical value $P_{c}\sim$ 8 kbar,  whereas the antiferromagnetic transition gets completely suppressed~\cite{Wu14}.
The superconducting transition temperature has the maximum $T_c\sim$ 2.2 K at about 1 GPa~\cite{Kotegawa14}. The mechanism of formation of Cooper pairs is still controversial since there are experimental clues pointing towards an unconventional order parameter, while some others  indicate a conventional superconducting mechanism. The first hypothesis is supported by nuclear quadrupole resonance, with the absence of a coherence effect in $1/T_1$ in the superconducting state~\cite{Kotegawa15}; the second conjecture is substantiated from muon spin rotation measurements, that show a phase separation scenario between superconductivity and magnetism, and a scaling law of $\rho_s$ as $\sim T_c^{3.2}$~\cite{Khasanov15}.

Considering the theoretical results, {\it ab-initio} calculations based on first-principles density functional theory give a good value for the magnetic moment, also suggesting that the CrAs is a weakly correlated material~\cite{Autieri17}.
Furthermore, an analysis based on a combination of the tight-binding approximation and the L\"{o}wdin down-folding procedure~\cite{Lowdin50,Andersen95,Noce99}, where the parameters entering the tight-binding matrix are the {\it ab-initio} derived overlap integrals for orbitals~\cite{Autieri17}, gives an accurate description of the electronic band structure and leads to results in good agreement with the experimental data~\cite{Autieri17b,Autieri18}. The features of the CrAs band structure, like the band degeneracies along high symmetry lines or the existence of energy gaps, reflect the underlying symmetries of the crystal and in particular the nonsymmorphic glide and screw symmetries, which involve half translation of a Bravais lattice. A recent study~\cite{Niu16} highlighted the role of such nonsymmorphic symmetries, which lead to nontrivial band crossings along specific high simmetry lines of the Brillouin zone (BZ), which are symmetry protected.

In this paper, we use the L\"{o}wdin down-folding procedure to study for the first time the effects of the intrinsic spin-orbit coupling (SOC) term on the electronic properties of the CrAs, assuming that the SOC may originate from Cr 3$d$ electrons as well as  from As 4$p$ ones. 
We show that, when the SOC is due to Cr electrons, one can get relevant modifications of the band structure and of the topology of the Fermi surface, along some high-symmetry lines of the BZ. On the other hand, no significant effects can be ascribed to the As electrons SOC.
In the specific, we demonstrate that the inclusion of the SOC allows for a removal of the band degeneracy due to the crystal symmetries, which only occurs along the $X-S$ and $T-Z$ paths of the BZ. Such constraint release of the number of  band crossings crucially determines a reconstruction of the Fermi surface along those high symmetry lines. For instance, it can reduce the number of electron pockets, by suppressing the small ones which are mostly affected by the SOC splitting. Moreover it acts by separating the residual pockets, which are no longer constrained by the crystal symmetry to be connected.

The paper is organized as follows: in the next section we will outline the procedure adopted, emphasizing how the SOC can be introduced within the down-folding approach; in Sec. III we will present our results confining the analysis to the modification of energy spectra and the Fermi surface when the SOC is considered. Sec. IV is devoted to the discussion and the comments of the results.

\section{Tight-binding model, L\"{o}wdin down-folding procedure and spin-orbit coupling}

\noindent The generic tight-binding model for the CrAs reads as:

\begin{equation}
\label{eqn:tightbinding}
H=\sum_{i,\alpha,\sigma}\epsilon^\alpha_{i}c^+_{i\alpha\sigma}c_{i\alpha\sigma}-\sum_{i,j,\alpha,\beta,\sigma}t^{\alpha\beta}_{ij}(c^+_{i\alpha\sigma}c_{j\beta\sigma}+h.c.)\, ,
\end{equation}\\

\noindent where the first term  describes the on-site energies of Cr and As orbitals, and the second one is related to the hoppings between the $d$ and/or $p$ orbitals of Cr and/or As atoms.
The sites are indexed by $i$ and $j$, the orbitals by $\alpha$ and $\beta$, while $\sigma$ is the spin.
As we have already said, in our model, the hopping parameters are the outcome of density functional theory calculations~\cite{Autieri17} for the overlap integrals between atoms and orbitals.
In order to calculate the energy spectra, we have to consider that each primitive cell contains four Cr and four As atoms, and correspondingly five $d$ and three $p$ orbitals. Thus, our tight-binding matrix Hamiltonian is a spin-degenerate $32\times32$ matrix, that can be divided in four sub-matrices related to the various hopping channels: 

\begin{equation}
	\label{eqn:matrix}
H=
\left(
\begin{array}{c|c}
H_{CrCr} & H_{CrAs} \\
\hline
H_{AsCr} & H_{AsAs}
\end{array}
\right)\, ,
\end{equation}

\noindent where $H_{CrCr}$ is a $20\times20$ matrix that contains the hoppings between the $d$ orbitals of the Cr atoms, $H_{AsAs}$ is a $12\times12$ one related to the As $p$ orbitals, while $H_{CrAs}$ and $H_{AsCr}$ describe the hopping terms between Cr and As atoms and vice-versa and are a $20\times12$ and a $12\times20$ sub-matrices, respectively.

Starting from {\it ab-initio} outcomes~\cite{Autieri17}, we consider only the nearest-neighbour hoppings, the second-nearest neighbour hoppings along the $x$-direction and the diagonal part of the second-nearest neighbour hoppings along the $y$ and $z$ direction, with the hopping values ranging from 80 meV to 1 eV~\cite{Autieri17b,Autieri18}.
To obtain the full band structure of the system we have to diagonalize a $32\times32$ matrix Hamiltonian. Nevertheless, we know from {\it ab-initio} calculations that the As bands are located above and below 2 eV from the Fermi level~\cite{Autieri17}, so that we can use the L\"{o}wdin down-folding procedure in order to get the low-energy effective bands, projecting out the As bands.
The effective Hamiltonian derived in this way is:

\begin{equation}\label{eqn:equationL}
\widetilde{H}_{CrCr}(\varepsilon)={H}_{CrCr}-{H}_{CrAs}\left({H}_{AsAs}-\varepsilon\mathbb{I}\right )^{-1}{H}_{AsCr}\, .
\end{equation} 

\noindent The result of the diagonalization of this $20\times20$ matrix has been previously used to evaluate the energy spectrum and the resistivity~\cite{Autieri17b} as well as  the evolution of the magnetic moment under pressure~\cite{Autieri18}. We would like to point out that obtaining an effective Cr-band model for CrAs, and other similar Cr-based superconductors, is highly non trivial~\cite{Edelmann17,Cuono18,Cuono18b}, and that even though the DFT results are available, they are not delivery in a form useful as the single-particle energy spectrum obtained from Eq.~(3).

\noindent In order to study the effects of SOC on the electronic properties of CrAs, we modify the diagonal terms of Hamiltonian of Eq.~(3)
including the single-particle on-site SOC term as follows~\cite{Friedel64}

\begin{equation}\label{eqn:equationSOCAsAs}
{H}^{SOC}_{AA}={H}_{AA}{\sigma}_{0}+ \frac{\lambda}{2}\left ( {L}_{x}{\sigma}_{x} + {L}_{y}{\sigma}_{y} + {L}_{z}{\sigma}_{z}\right )\, ,
\end{equation} 

\noindent where ${\sigma}_{0}$ is the unit matrix in the space of spins, ${\sigma}_{x}$, ${\sigma}_{y}$ and ${\sigma}_{z}$ are the Pauli matrices, ${\lambda}$ is the SOC constant and ${A}$=As, Cr depending on the origin for the SOC we want to analyze.
In the next section, we will analyze the results of our calculations, considering first the effects on the bands of a SOC due to As and Cr atoms separately, and then simultaneously.

\section{SOC effects on the band structure and Fermi surface}

\begin{figure} [t]
	\centering
	\includegraphics[width=8cm]{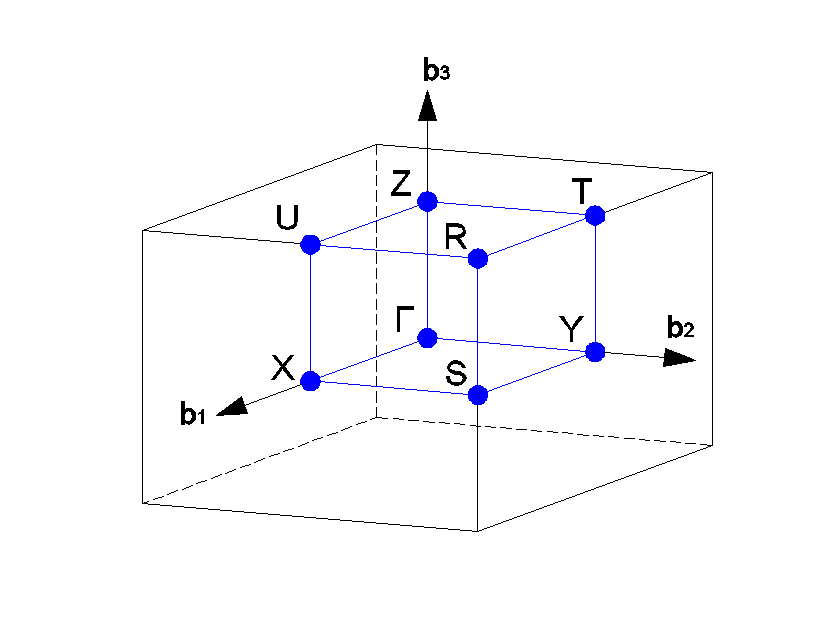}
	\caption{High-symmetry path in the orthorombic Brillouin zone chosen according to the notation quoted in Ref.~\cite{Setyawan10}}
	\label{fig:kspace}
\end{figure}

\noindent In this section, we analyze the effect of the inclusion of the SOC on the electronic band structure of CrAs. We assume the SOC terms on the Cr and As ions are relevant for describing the electronic structure, due to the significant degree of covalency between the 3d and 4p orbitals. In the following, the present calculations are shown in comparison with the outcome of our previous tight-binding analysis, where the SOC has not been included~\cite{Autieri17b}. The high-symmetry points along which we will plot the band structure have been chosen according to the notation quoted in Ref.~\cite{Setyawan10}, and the path is represented in Fig.~\ref{fig:kspace}.

In absence of SOC, the Hamiltonian possesses several symmetry properties, which are responsible for the characteristic features of the band structure~\cite{Autieri17b}.  First of all, it is symmetric upon the action of the time-reversal operator T, which generates the time-symmetry transformation. Moreover, the system is inversion symmetric via the unitary operator P that inverts the sign of the in-plane spatial coordinates. Then, the PT combination allows to have an antiunitary symmetry that is local in momentum space and implies that the energy eigenstates realize bands which are twofold degenerate at any $k$ vector in the Brillouin zone. This Kramers degeneracy sets to 20 the maximal number of bands, each with a 2-fold degeneracy. Apart from the time and inversion symmetry, the system has additional nonsymmorphic symmetries of the crystal structure, which have been recently discussed in relation with the occurrence of a 4-fold degeneracy along specific lines of the BZ, as well as to nontrivial band crossings at high symmetry points of the Brillouin zone, which are symmetry protected~\cite{Niu16}. In order to analyze how the SOC may affect the features of the band structure, we first observe that SOC doesn't break the time-reversal symmetry, so that the Kramers degeneracy is preserved.

\begin{figure} [t]
	\centering
	\includegraphics[width=8cm]{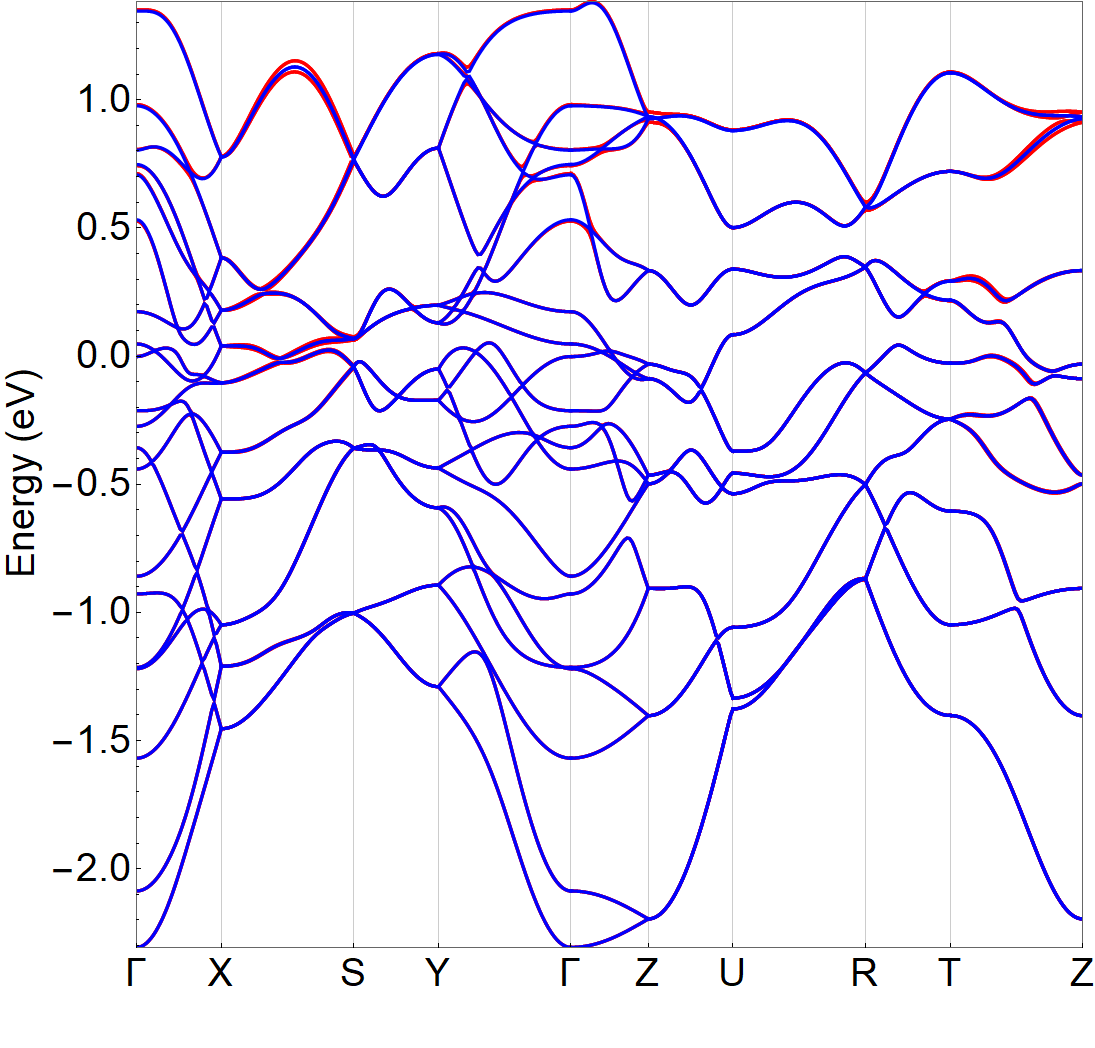}
	\caption{Comparison between the low-energy bands of CrAs without SOC (blue) and the modified bands with the action of the SOC on the 4$p$ orbitals of the As atoms (red). The value of the SOC constant used is ${\lambda}_{As}$=0.3 eV. The Fermi level is set at zero energy.}
	\label{fig:bandeas}
\end{figure}

We start by considering the effects of the SOC as due to the As electrons only. {\it Ab-initio} calculations give ${\lambda}_{As}$=0.164 eV~\cite{Autieri17}; however, since this coupling constant gives rise to negligible effects, we choose to emphasize the value of ${\lambda}_{As}$ by assuming ${\lambda}_{As}$=0.3 eV, in order to highlight the regions of the energy spectrum that are more affected by the As SOC term. In Fig.~\ref{fig:bandeas}, we show the comparison between the low-energy bands of CrAs without SOC (blue lines) and those obtained considering the effect of the SOC on the As degrees of freedom (red lines). We notice that the arsenic SOC contribution to the low energy bands of CrAs is rather weak, and it mainly interests the bands far from the Fermi level, even when a larger value of ${\lambda}_{As}$ is assumed.
Furthermore, we observe that the 4-fold degeneracy is removed along the $X-S$ and $T-Z$ paths, the splitting of the bands along these directions being of the order of $10^{-2}$ eV. 

\begin{figure} [h]
	\centering
	\includegraphics[width=8cm]{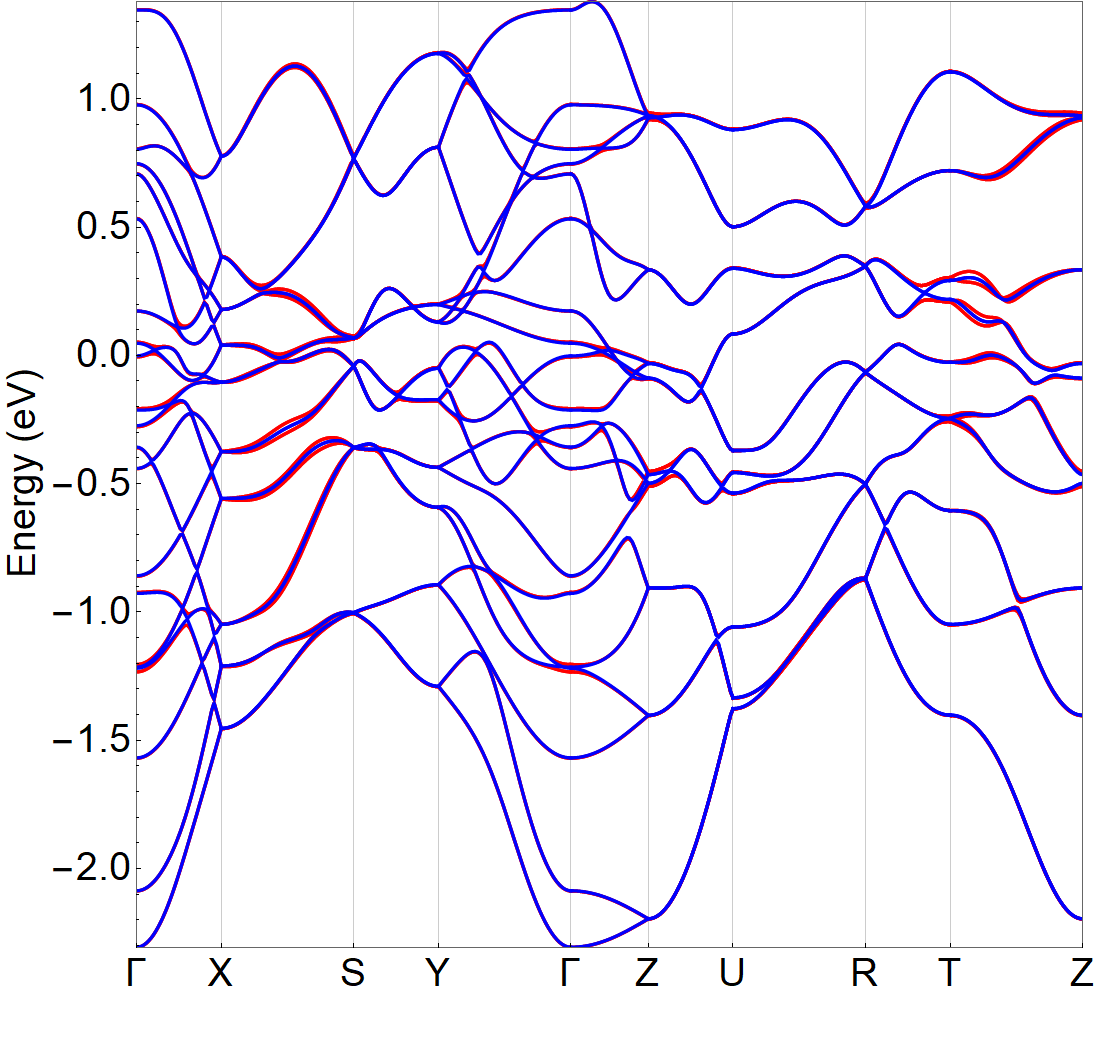}
	\caption{Comparison between the low-energy bands of CrAs without SOC (blue) and the modified bands after the action of the SOC on the 3$d$ orbitals of the Cr atoms (red). The value of the SOC constant used is ${\lambda}_{Cr}$=0.033 eV. The Fermi level is set at zero energy.}
	\label{fig:bandecr}
\end{figure}

\noindent As next step, we have considered the SOC contribution on Cr ions, by modifying the Cr-Cr part of the Hamiltonian of Eq.~(\ref{eqn:matrix}) and then performing the L\"{o}wdin procedure~\cite{Autieri17b}, to down-fold the $H_{AsAs}$ matrix. In this way, we obtain the band structure via the resolution of a corresponding $20\times20$ effective Hamiltonian, which is expressed only in terms of the Cr and where the sub-dominat low-lying As degrees of freedom have been projected out.
In this case, according to {\it ab-initio} calculations, we assume ${\lambda}_{Cr}$=0.033 eV~\cite{Autieri17}. We notice that this value of the SOC parameter is in good agreement with the
value (0.034 eV) found in another Cr anisotropic compound with monoclinic space group~\cite{Autieri14b}.\\
The comparison between the bands without SOC and the bands where the Cr SOC is considered is shown in Fig.~\ref{fig:bandecr}.

\begin{figure} [h]
	\centering
	\includegraphics[width=8cm]{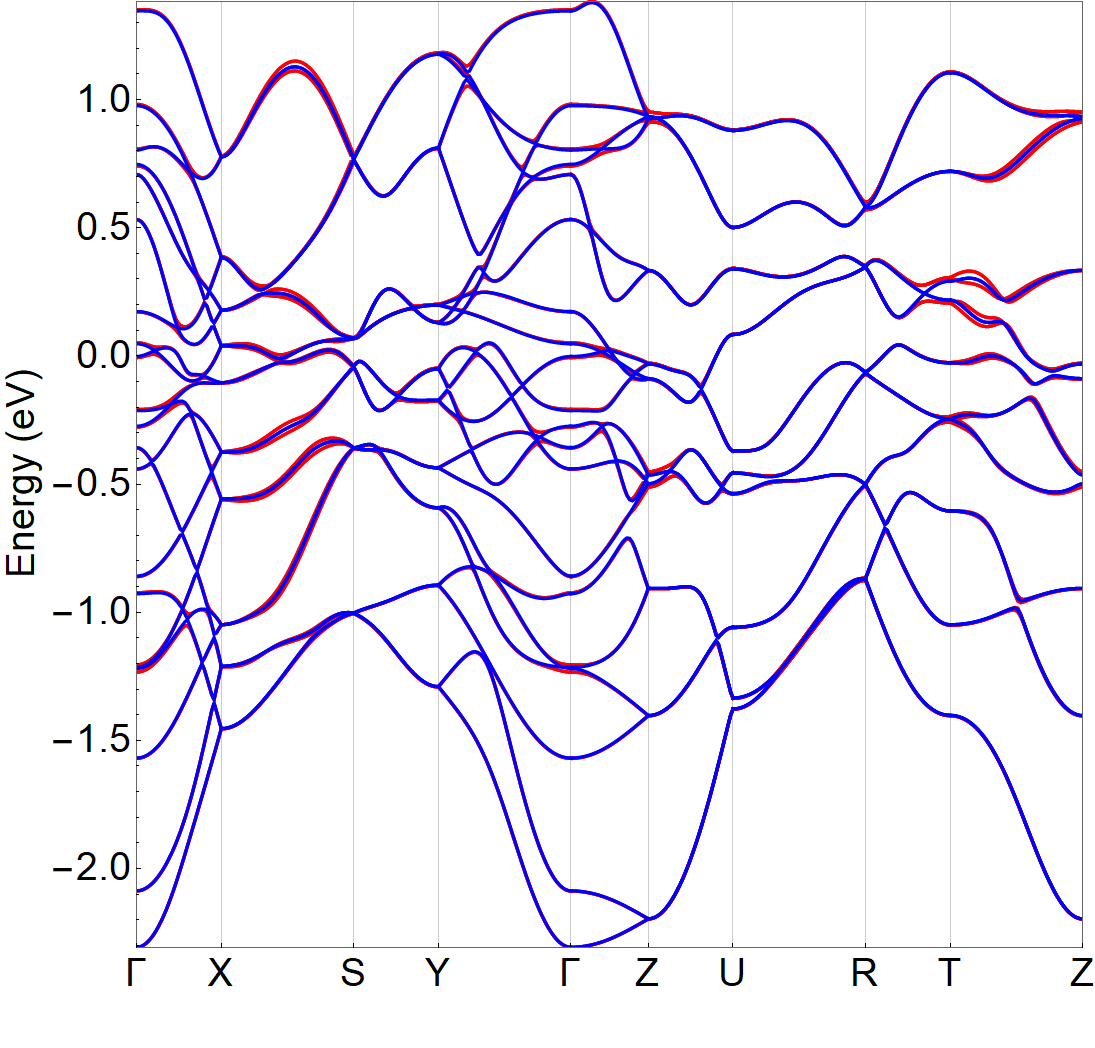}
	\caption{Comparison between the low-energy bands of CrAs without SOC (blue) and the modified bands after the action of the SOC both on the Cr and on the As part of the Hamiltonian (red). The values of the SOC constants used are ${\lambda}_{Cr}$=0.033 eV and ${\lambda}_{As}$=0.164 eV. The Fermi level is set at zero energy.}
	\label{fig:bandetot}
\end{figure}

\noindent From an inspection to this figure, we infer that the Cr SOC affects the bands near the Fermi level, implying that this effect may be relevant on the transport properties of the CrAs.
We point out that this change underlines the fact that the bands close to the Fermi level are mainly due to the chromium orbitals. Also in this case, the SOC causes a partial removal of the degeneracy, from a 4-fold to a 2-fold, along the $X-S$ and $T-Z$ high-symmetry lines of the orthorhombic BZ.\\ 
\noindent Finally, in Fig.~\ref{fig:bandetot} we plot the modification to the bands when the SOC acts both on the Cr and on the As part of the Hamiltonian. The overall effect is a degeneracy removal that corresponds to an energy splitting of the order of $10^{-2}$ eV, taking place on the above mentioned high-symmetry lines of the BZ. \\

In Table 1, we summarize the change in the bands degeneracy induced by the SOC, along each line of the orthorombic BZ. As one can see, the inclusion of the SOC leads to a partial removal from a 4-fold to the 2-fold Kramers degeneracy in the single-particle electron states, which only occur along the high-symmetry lines where $k_x=\pi/a$ and $k_z=0$, while varying $k_y$ ($X-S$), and $k_x=0$ and $k_z=\pi/c$, while varying $k_y$ ($T-Z$). The non-trivial interplay between the SOC and the non-symmorphic crystal symmetries along the $y$ axis, which underlies this selective degeneracy removal is left for future investigations.

\begin{table} [] 
	\noindent \begin{flushleft}
		\begin{tabular}{|c||c|c|c||c|c|c|} 
			\hline 
			High-symmetry line & $k_x$ & $k_y$ & $k_z$ & TB & L\"{o}wdin & L\"{o}wdin with SOC\\ \hline
			$\Gamma$-X & $k_x$  & 0 & 0 & 2 & 2 & 2 \\ \hline
			X-S & $\frac{\pi}{a}$ & $k_y$ & 0 & 4 & 4 & 2 \\ \hline
			S-Y & $k_x$ & $\frac{\pi}{b}$ & 0 & 4 & 4 & 4 \\ \hline
			Y-$\Gamma$ & 0 & $k_y$ & 0 & 2 & 2 & 2 \\ \hline
			$\Gamma$-Z & 0 & 0 & $k_z$ & 2 & 2 & 2 \\ \hline
			Z-U & $k_x$ & 0 & $\frac{\pi}{c}$ & 4 & 4 & 4 \\ \hline
			U-R & $\frac{\pi}{a}$ & $k_y$ & $\frac{\pi}{c}$ & 4 & 4 & 4 \\ \hline
			R-T & $k_x$ & $\frac{\pi}{b}$ & $\frac{\pi}{c}$ & 4 & 4 & 4 \\ \hline
			T-Z & 0 & $k_y$ & $\frac{\pi}{c}$ & 4 & 4 & 2 \\ \hline
		\end{tabular}
		\end{flushleft}
	\caption{Evolution of the bands degeneracy along high symmetry lines in the BZ, with the introduction of the SOC term. Along each line, one component of $\bm{k}$ varies and the other two are kept fixed. We indicate in the first column the high-symmetry lines of the Brillouin zone; in the second, third and fourth columns the values assumed by the components of $\bm{k}$; in the fifth column, we report as a comparison, the degeneracy value obtained in the pure TB procedure; the sixth and the seventh columns show the band degeneracy without and with the inclusion of the SOC. \label{tab1}}
\end{table}

\begin{figure} [h]
	\centering
	\includegraphics[width=8cm]{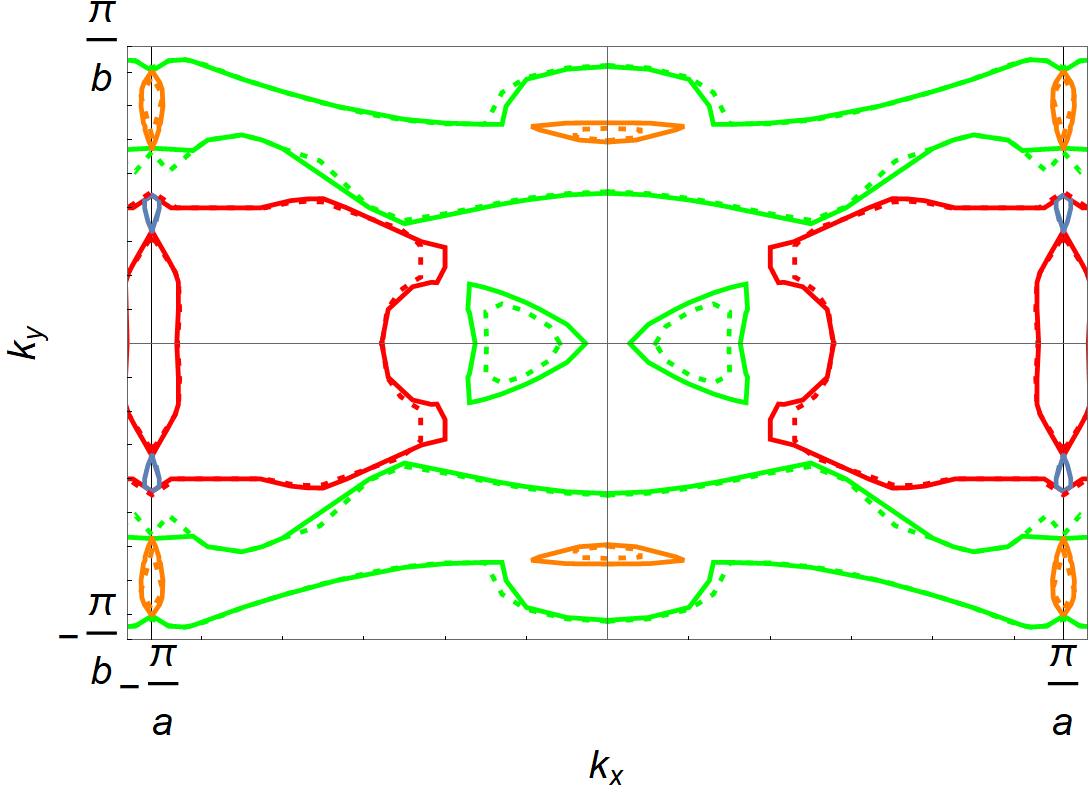}
	\caption{Spin-orbit coupling effects on the in-plane Fermi surface at $k_z=0$ of CrAs, obtained  for ${\lambda}_{Cr}$=0.033 eV and ${\lambda}_{As}$=0.164 eV. The different colors indicate the different bands that cross the Fermi level while the dotted line represents the modified Fermi surface by the SOC.
	}
	\label{fig:fermi}
\end{figure}

Since the SOC is acting by splitting the 4-fold band degeneracy, it is plausible to expect that the Fermi surface topology will be strictly affected, by disconnecting the pockets that were constrained by the crystal symmetry to cross each other along those high-symmetry lines. In Fig.\ref{fig:fermi}, we show the representative case of the in-plane Fermi surface at $k_z=0$, obtained when a SOC term acting both on the Cr and on the As atoms is considered (dashed lines), as compared to the one obtained without the inclusion of the SOC (solid lines). The {\it ab-initio} values for the constants, namely ${\lambda}_{Cr}$=0.033 eV and ${\lambda}_{As}$=0.164 eV, have been used.

\noindent In absence of SOC, the Fermi surface consists of four sheets, and is characterized by the presence of four pockets along $X-S$, which are two at a time degenerate at four specific $k_y$ values. The energy splitting induced by the SOC along this path has two main effects. First of all, by lowering one of the bands below the Fermi level, it causes the removal of the small electron pockets located at $(k_x=\pi/a, k_y\simeq 0.5$) (blue pockets). Moreover, the residual pockets are no longer constrained by the crystal symmetry to be connected, so they get separated and cut the $X-S$ path at eight different $k_y$ values.
We would also like to point out that the modification to the Fermi surface is essentially due to the Cr part of the SOC, once again supporting the conjecture that the bands near the Fermi level are mainly due to the Cr orbitals. 

\section{Conclusions and remarks}

We have used a combined tight-binding-L\"{o}wdin down-folding procedure to get the low-energy band structure of CrAs, and to evaluate the role played by the SOC on the electronic properties of this material. The main effect of the inclusion of the SOC term in our model Hamiltonian is the removal of the band degeneracy along the $X-S$ and $T-Z$ lines of the orthorhombic Brillouin zone. The SOC splitting of the bands along these directions is of the order of $10^{-2}$ eV. We demonstrate that, under the action of the SOC term, the topology of the Fermi Surface along those high symmetry lines is also modified. In particular, such energy splitting is substantial as it affects the small electron pockets located along $X-S$ and $T-Z$ paths, which get suppressed once the constraint of the band 4-fold degeneracy is released by the introduction of the SOC. We are able to ascribe to the 3$d$ orbitals of the chromium this effect, further supporting the conjecture that the spectral weight of the $d$ band near the Fermi level is larger than the $p$-band corresponding one.
We stress that band crossings protected by non-symmorphic symmetry have attracted a lot of interest since they may give rise to interesting features involving unconventional dispersions, novel topological response phenomena and unusual transport in external magnetic field. A detailed study of the symmetries of the band spectrum of the CrAs when the SOC is included deserves a special attention; this job is currently underway.\\ 
Finally, we underline the importance of this kind of exploration since it is well known that SOC may affect the superconducting properties, as it has been shown to induce an enhancement of the coupling constant, both softening the phonon spectrum and/or increasing the electron-phonon coupling matrix elements~\cite{shao16,heid00,smirnov18}. An investigation of this issue for CrAs is planned in the near future.
Recently, the discovery of the new superconductor WP has been published on arXive, see Ref.~\cite{Liu19}.
The stoichiometry, the filling, group symmetry, and the spin-orbital properties are the same of our paper though the elements differ. The spin-orbit effects in WP are expected to be more relevant than in the CrAs since the $d$-element is heavier.

\section*{Acknowledgments} 
The work is supported by the Foundation for Polish Science through the IRA Programme co-financed by EU within SG OP.

\end{document}